\documentclass[a4paper]{article}

\usepackage{INTERSPEECH2022}
\usepackage{multirow}
\usepackage{makecell}
\usepackage{cite}
\usepackage{url}

\title{Locality Matters: A Locality-Biased Linear Attention\\ for Automatic Speech Recognition}

\name{Jingyu Sun$^{1,2}$, Guiping Zhong$^{1}$, Dinghao Zhou$^{1}$, Baoxiang Li$^{1}$, Yiran Zhong$^{1}$}
\address{
  $^1$SenseTime Research\\
  $^2$Beijing University of Posts and Telecommunications}
\email{\{sunjingyu,zhongguiping,zhoudinghao,libaoxiang,zhongyiran\}@sensetime.com}
\begin{document}
\maketitle

\begin{abstract}
Conformer has shown a great success in automatic speech recognition (ASR) on many public benchmarks. One of its crucial drawbacks is the quadratic time-space complexity with respect to the input sequence length, which prohibits the model to scale-up as well as process longer input audio sequences. To solve this issue, numerous linear attention methods have been proposed. However, these methods often have limited performance on ASR as they treat tokens equally in modeling, neglecting the fact that the neighbouring tokens are often more connected than the distanced tokens. In this paper, we take this fact into account and propose a new locality-biased linear attention for Conformer. It not only achieves higher accuracy than the vanilla Conformer, but also enjoys linear space-time computational complexity. To be specific, we replace the softmax attention with a locality-biased linear attention (LBLA) mechanism in Conformer blocks. The LBLA contains a kernel function to ensure the linear complexities and a cosine reweighing matrix to impose more weights on neighbouring tokens. 
Extensive experiments on the LibriSpeech corpus show that by introducing this locality bias to the Conformer, our method achieves a lower word error rate with more than 22\% inference speed.

\end{abstract}

\section{Introduction}
\label{introduction}

ASR is a long-lasting problem in audio processing and has been extensively studied for decades. Recently, end-to-end deep neural network based methods are becoming one of the most promising approaches and have dominated numerous ASR leader boards. An end-to-end ASR model often integrates acoustic model, pronunciation model and language model coherently. There are several widely used structures for end-to-end ASR modelling such as connectionist temporal classification (CTC)~\cite{dahl2011context,mohri2008speech,graves2006connectionist}, attention-based encoder–decoder model (AED)~\cite{miao2015eesen,chorowski2014end,chorowski2015attention,zhang2017very}, recurrent neutral network transducer (RNN-T)~\cite{graves2012sequence,graves2013speech,rao2017exploring} and  Transformer~\cite{vaswani2017attention}. In particular, Conformer~\cite{gulati2020conformer} is a variant of Transformer and the most advanced structure for ASR due to its high performance.


Attention is the key component in Conformer which aims to capture global dependencies. Compared with conventional encoders such as the long short-term memory (LSTM)~\cite{graves2013speech} which implicitly encodes temporal information in hidden layers, the attention-based encoders explicitly store the global temporal dependencies through a relevance weighted matrix and use it to weight the input sequence. Although previous methods demonstrate the effectiveness of the attention mechanism, it suffers two crucial drawbacks: 1). It is less effective to capture local dependencies; 2). It requires quadratic space-time computational complexity with respect to the sequence length. Numerous approaches have been proposed to address these limitations.

To emphasize the local dependencies, ContextNet~\cite{han2020contextnet} proposes using convolution module to capture local dependencies and leveraging a squeeze-and-excitation module to summarize global context from input sequence. Conformer integrates the convolution module and self-attention network (SAN) module in a transformer block in a sequential manner to augment local dependencies and obtains the state-of-the-art performance. In this paper, we propose a new locality-biased attention to blend local dependencies in a SAN module as well and achieves a higher performance.

For the quadratic computational complexity, a common solution in ASR is to restrict the input speech length or truncate the long form speech into several smaller speeches. However, the model's performance may degrade due to the lack of long-range information dependencies. This solution does not touch the essence of the limitation as the network still suffers the quadratic complexity. Efficient transformers are then proposed to reduce the quadratic space-time complexities directly. Sparse transformers often use local windows~\cite{beltagy2020longformer}, learnable sparse attention matrix~\cite{tay2020sparse},  locality-sensitive hashing attention mechanism~\cite{kitaev2020reformer}, and Kullback-Leibler divergence based sparsity measurement~\cite{wang2021efficient} to skip some attention weight's computation and thus reduce the computational complexities. Linear transformers, on the other hand, reduce the complexities to linear with some kernel tricks. For example, the Gaussian-kernalized self-attention~\cite{kashiwagi2021gaussian} uses a variant of the Gaussian kernel to represent attention weight and replace dot-product self-attention.
Random Feature Attention (RFA)~\cite{peng2021random} proposes a linear time and space attention that uses random feature methods to approximate the softmax function and applies in Transformer. Performers~\cite{choromanski2020rethinking} estimates vanilla attention with comparable accuracy using a novel fast attention via positive orthogonal random features approach, which has linear space-time complexities. Since they involved softmax approximation in the attention process, and these errors may accumulated through layers, these methods often suffer from degenerated performances. For the ASR task, as it is trying to map the continuous speech features into discrete transcripts, it is more sensitive to these approximation errors and will cause severer performance drop.  

In this paper, we solve these two drawbacks with one stone: a new locality-biased linear attention that not only ensures neighbouring tokens to have a higher weights than the distanced tokens but also enjoys linear space-time computational complexity. Our LBLA consists of two components: a kernel function that allows us to use kernel tricks to reduce the computational complexity to linear, and a locality-biased re-weighting mechanism to impose the locality bias. It is worth noting that not all locality biased re-weighting matrix can be applied here as it also needs to satisfy the linear attention requirements, so that we can use the kernel tricks on the re-weighting matrix as well. The experiments are conducted on the LibriSpeech corpus~\cite{panayotov2015librispeech}. They show that compared with the vanilla Conformer model, the proposed method increases the inference speed by 6\%$\sim$22\% and obtains the comparable word error rate (WER) performance. After finetuning with model initialization, the proposed method obtains WER reduced by relatively 2.4\%$\sim$5.7\%.

The paper is organized as follows: Section \ref{rnnt} gives a brief introduction about conformer encoder architecture. The proposed conformer-based LBLA model is described in Section \ref{sec_hmm_free}, followed by experiments and discussions in Section \ref{src_exp}.

\section{Preliminaries}
\label{rnnt}
\subsection{Softmax Attention}
End-to-end ASR model maps the arbitrary input feature sequence $X$=$\{x_{1},x_{2},..x_{t}\}$ of length $T$ into output sequence $Y$=$\{y_{1},y_{2},...y_{u}\}$ of length $U$ directly. Usually, the length of label $U$ is much smaller than length of speech frames $T$. Transformer is proposed in~\cite{vaswani2017attention}, which is a stack of transformer layer consists of self-attention network (SAN), feed forward network (FFN), layernorm and residual connection. Self-attention mechanism in SAN can be represented below:
\begin{equation}
\label{eq: attention}
{O} =\text { Attention }(Q, K, V)= \operatorname{Softmax}\left(\frac{Q K^{T}}{\sqrt{d_{k}}}\right)V
\end{equation}
where $d_{k}$ denotes the dimension of $K$, $\frac{1}{\sqrt{d_{k}}}$ is scaling factor, $Q$, $K$ and $V$ refer to the query, key and value that are projected to high-level vectors using parameter matrices $W_{q}$,$W_{k}$ and $W_{v}$:
\begin{equation}
    Q = XW_{q},K = XW_{k},V = XW_{v}
\end{equation}
Multi-head attention (MHA) mechanism also proposed in~\cite{vaswani2017attention} projects query, key and value to several separately linear vector subspace. Then all of value embedding concatenated and project again with $W_{o}$:
\begin{equation}
\text { $head_{i}$ }=\operatorname{Attention}(Q_{i}W_{k}^{i},K_{i}W_{k}^{i},V_{i}W_{k}^{i})
\end{equation}
\begin{equation}
{O} =\text { MHA }(Q, K, V)=\operatorname{concat}({head_{1},..., head_{h}})W_{o}
\end{equation}
Where $h$ is the number of heads and $W_{q}^{i}$,$W_{k}^{i}$ and $W_{v}^{i}$ 
are parameter matrices for i-th attention head.

\subsection{Linear Attention}
As shown in Eq.~\ref{eq: attention}, query  $Q \!\in\! \mathbb{R}^{T\times d}$ and  key $K \!\in\! \mathbb{R}^{T\times d}$ are utilized for the attention weight matrix to represent the proximity between them, where $T$ is the length of input sequence and $d$ is the embedding length. Self-attention mechanism calculates attention weight matrix $QK^{T}\!\in\! \mathbb{R}^{T\times T}$ followed by softmax function that is applied to obtain normalization.

Since the complexity of calculating attention weight matrix is $\mathcal{O}(T^{2})$, which has quadratic time and memory bottleneck with respect to the input sequence length. In order to address this issue, we approximate the attention weight matrix and rewrite self-attention function as: 

\begin{equation}
 \quad {O}_{i}=\sum_{j} \frac{{exp}\left(Q_{i}K_{j}^T\right) V_{j}}{\sum_{j} {exp}\left(Q_{i}K_{j}^T\right)},
\end{equation}
\begin{equation}
{O}=\left[{O}_{1}, \ldots,{O}_{T}\right]^{T}
\end{equation}
Denote operator $\exp(Q_{i}K_{j}^T)$ as the proximity between i-th query and j-th key, the proximity measure function $\mathcal{P}(\cdot)$ can be defined as:
\begin{equation}
 \quad  \mathcal{P}(Q_{i},K_{j})=\operatorname{exp}(Q_{i}K_{j}^T)
\end{equation}
Then self-attention can be rewrite as: 
\begin{equation}
 \quad {O}_{i}=\sum_{j} \frac{\mathcal{P}\left(Q_{i}, K_{j}\right) V_{j}}{\sum_{j} \mathcal{P}\left(Q_{i}, K_{j}\right)}
\end{equation}
The proximity measure function $\mathcal{P}(\cdot)$ need to be calculated $T^2$ times. To maintain a linear computation complexity, we adopt a decomposable proximity function below:
\begin{equation}
\mathcal{P}(Q_{i}, K_{j})=\psi(Q_{i})\psi(K_{j})^T
\label{eq: decomp}
\end{equation}
where $\psi(\cdot)$ is a kernel function that maps the queries and keys to hidden representations. Consequently, we could rewrite self-attention in the form of kernel functions as:
\begin{equation}
 \quad {O}_{i}=\sum_{j}\frac{\psi(Q_{i})(\psi(K_{j})^TV_{j})}{\sum_{j} \psi(Q_{i})\psi(K_{j})^T}
 \label{eq: trick}
\end{equation}
Instead of explicitly calculating the attention weight matrix $QK^{T}$$\in$$\mathbb{R}^{T\times T}$, key and value are calculated first as  $\psi(K_{j})^TV_{j}$$\in$$\mathbb{R}^{d\times d}$. Note that the dimension of head $d$ is always much smaller than the input speech sequence length of $T$. By using this trick, the computational complexity of calculating matrix can be reduce to $\mathcal{O}(Td^2)\approx \mathcal{O}(T)$ when $T\gg d$.


\section{Locality-Biased Linear Attention}
\label{sec_hmm_free}
In this section, we first provide a detailed description of the two components in our LBLA module, \emph{i.e.,} the kernel functions and the locality-biased re-weighting matrix, then illustrate how to integrate the LBLA module to a Conformer block.


\subsection{The Kernel Function}
To linearize the attention, the kernel functions have to satisfy the Eq.~\ref{eq: decomp}, then we can apply the kernel trick in Eq.~\ref{eq: trick} to reduce the computational complexities. How to find the right kernel function for the ASR task is a non-trivial task. Previous methods in natural language processing~\cite{katharopoulos2020transformers, zhen2021cosformer} suggest that the kernel function needs to project the tokens to a non-negative space such that the negatively-correlated information can be excluded in contextual information aggregation. Weak-attention~\cite{shi2020weak} suggests that ASR will likely benefit from sparse and localized attention weight. Following these suggestions, we explore three commonly used non-negative activation functions as our kernel functions: the ReLU function, the Exponential function, and the Sigmoid function. We defined these functions as below:
\begin{align}
 \psi(x)&=\operatorname{ReLU}(x), \\
 \psi(x)_{exp}&=\operatorname{exp}(x),\\
 \psi(x)_{sig}&=\operatorname{Sigmoid}(x).
\end{align}
We use dot product as our proximity measure function $\mathcal{P}(\cdot)$ so that the requirement of Eq.~\ref{eq: decomp} fulfilled.

In the experiment section, we extensively ablate these selections and find that the Sigmoid one achieves the most accurate results. Therefore, in our LBLA, we use the Sigmoid function as our kernel function.


\subsection{Locality-biased Re-weighting Mechanism}
The Eq.~\ref{eq: decomp} computes the proximity measurements without any re-weighting mechanism. In Eq.~\ref{eq: reweight}, we show a locality-biased proximity measurement with a re-weighting matrix $\omega(\cdot)$.
\begin{equation}
\mathcal{P}_{local}(Q_{i}, K_{j})=\psi(Q_{i})\psi(K_{j})^T\omega(i-j)
\label{eq: reweight}
\end{equation}
where the $\omega(\cdot)$ needs to satisfy the Eq.~\ref{eq: linear} for linearization.
\begin{equation}
\textstyle
\omega(i-j) = \sum_1^n f(i)g(j)
\label{eq: linear}
\end{equation}
where $f(\cdot), g(\cdot)$ are decomposed functions of $\omega$ and $n$ is the number of decomposed terms. 

Not many re-weighting matrices meet this requirement. Here, we use a cosine re-weighting matrix which can be decomposed with the Ptolemy's theorem as below:
\begin{equation}
 \begin{scriptsize}
 \begin{aligned}
 \omega_{\operatorname{cos}}(i-j) &=
\operatorname{cos}(\frac{\pi}{2}\!\times\!\frac{i\!-\!j}{T})\\ &= \cos (\frac{\pi i}{2 T})\cos (\frac{\pi j}{2 T})\!+\!\sin (\frac{\pi i}{2 T}) \sin (\frac{\pi j}{2 T})
\end{aligned}
 \end{scriptsize}
\end{equation}

\begin{figure}[t]
	\centering
	\includegraphics[width=0.9\linewidth,height=3cm]{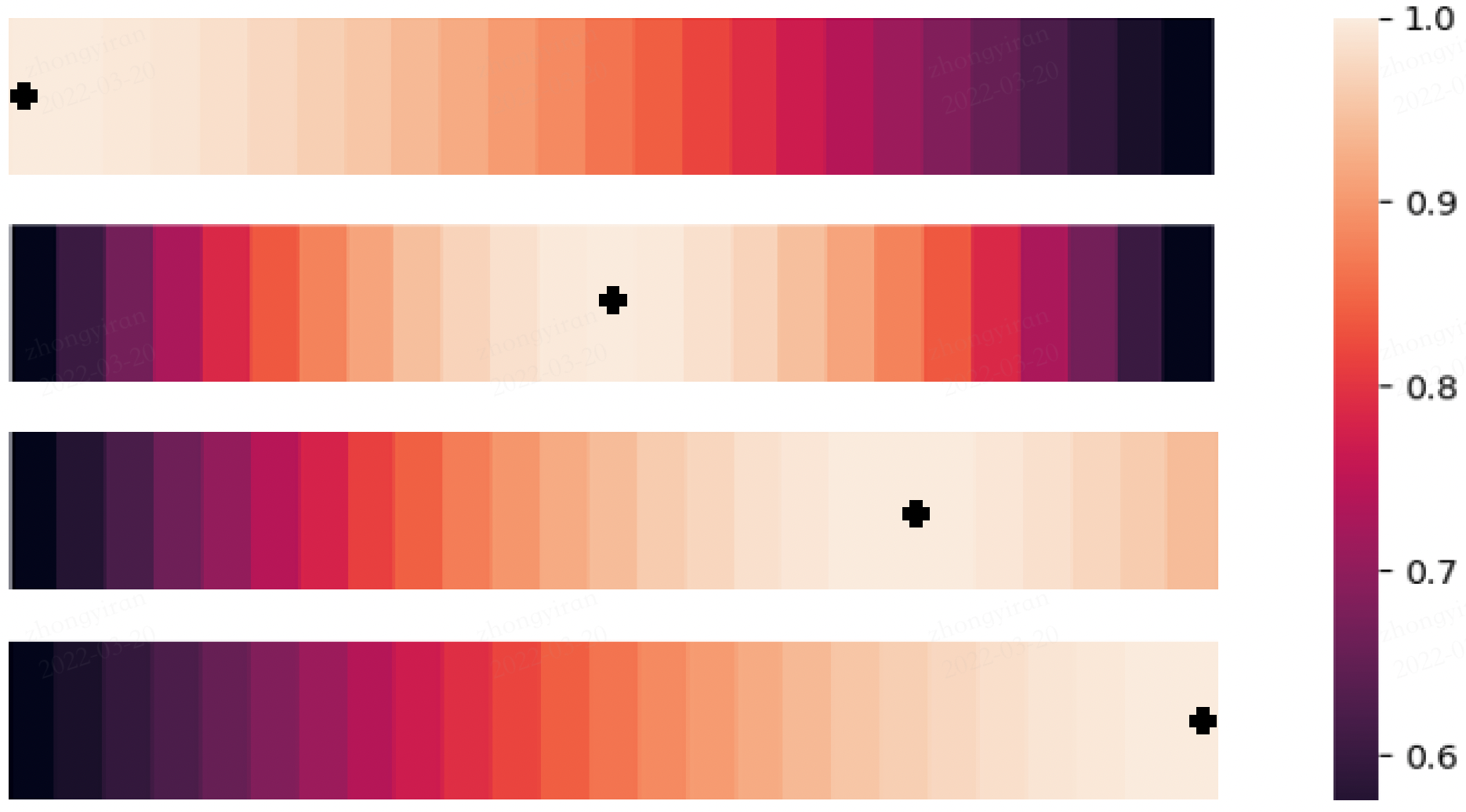}
	\caption{Visualization of the locality-biased re-weighting patterns. The black cross denotes the query position and the mask denotes the corresponding locality weights.}
	\label{fig:locality}
\end{figure}

Figure~\ref{fig:locality} illustrates our locality-biased re-weighting matrix, where the neighbouring tokens are assigned with higher weights and the distanced ones are with lower weights.

We now achieve the overall expression of our LBLA:
\begin{equation}
\textstyle
{O}= \frac{\mathcal{P}_{local}(Q, K) V}{\sum \mathcal{P}_{local}(Q, K)}= \frac{Q^{\cos }(K^{\cos } V)+Q^{\sin }(K^{\sin } V)}{\sum Q^{\cos }K^{\cos } +\sum Q^{\cos }K^{\cos } }
\end{equation}
where \scalebox{0.9}{$Q_{i}^{cos}\!=\!\widetilde{Q}_{i}\cos (\frac{\pi i}{2 M}), Q_{i}^{sin}\!=\!\widetilde{Q}_{i}\sin (\frac{\pi i}{2 M}),K_{j}^{cos}\!=\!\widetilde{K}_{j}\cos (\frac{\pi j}{2 M})$}, \scalebox{0.9}{$K_{j}^{sin}\!=\!\widetilde{K}_{j}\sin (\frac{\pi j}{2 M})$, $\widetilde{Q}_{i}=\psi(Q_{i}), \widetilde{K}_{j}=\psi(K_{j})$}.
Compared with vanilla softmax attention, our LBLA has a linear computational complexity with locality-biased re-weighting mechanism.

\subsection{Locality-biased Conformer Block}

\begin{figure}[t]
	\centering
	\vspace{-0mm}
	\includegraphics[width=\linewidth]{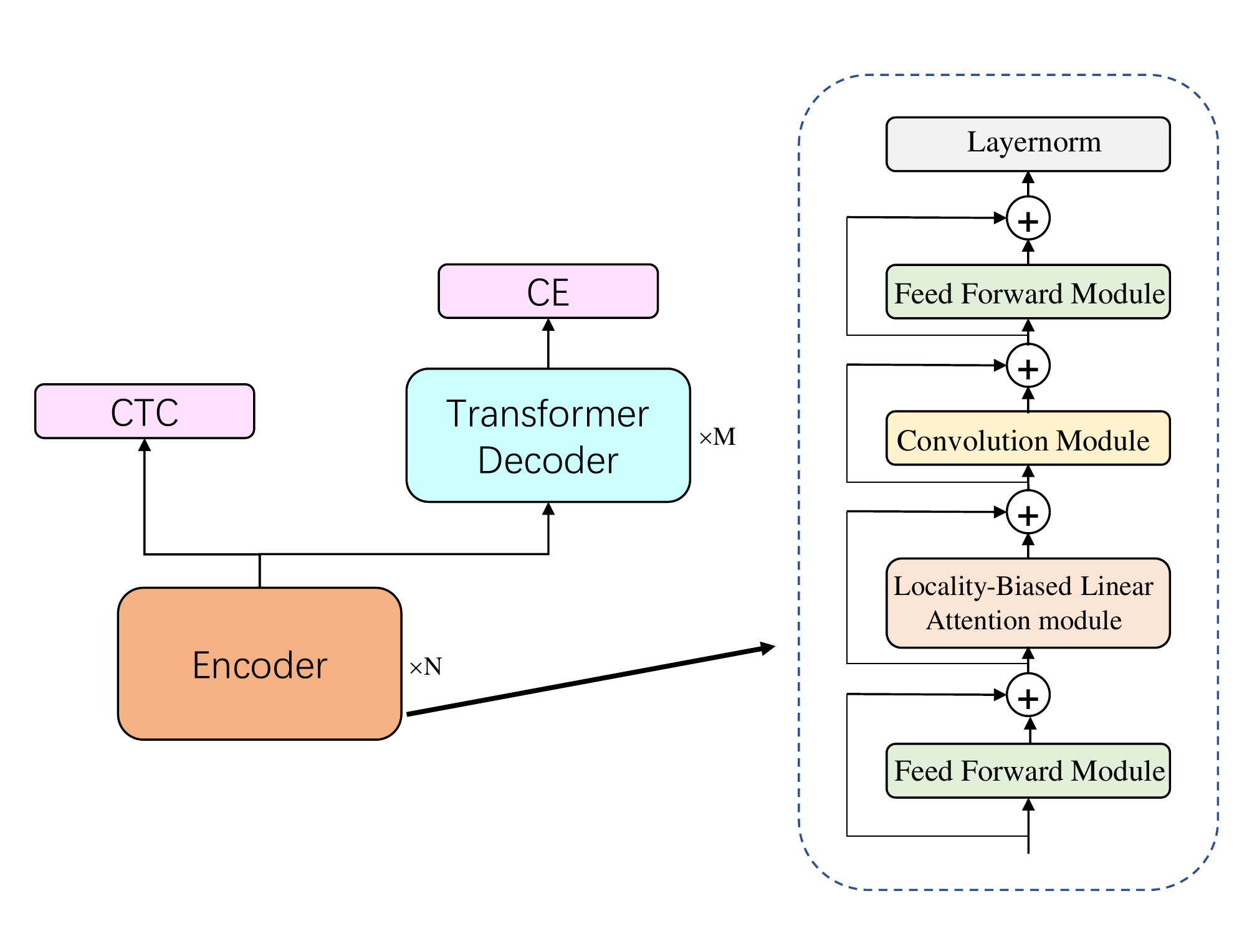}
	\vspace{-8mm}
	\caption{The structrue of LBLA based conformer model. The encoder contains locality-biased self-attention module, macaron-like feed-forward module and convolution module, followed by a post layernorm}
	\label{fig:architecture}
\end{figure}

Conformer has obtained state-of-the-art performance in ASR~\cite{gulati2020conformer,guo2021recent,li2021better,ng2021pushing,chung2021w2v,zhang2021bigssl}. As a variant of Transformer, Conformer uses convolution module to augment local dependencies. A single vanilla conformer block consists of convolution module, multi-head self-attention module and two independent feed forward networks. In our locality-biased Conformer block, we replace the multi-head self-attention module with our LBLA module and keep the rest modules unchanged. As illustrated in Figure~\ref{fig:architecture}, our block starts with a FFN, then the LBLA module to capture locality-biased global context followed by a convolution module that captures local context, another FFN is deployed as the last module. Especially, FFN is employed half-step residual weights in Conformer. The convolution module consists of pointwise convolution and gated linear unit, followed by one dimension convolution module and batchnorm, the last in swish activation function and another pointwise convolution.

\section{Experiments}
\label{src_exp}
\subsection{Experimental Setup}
The experiments are conducted using LibriSpeech corpus, which has 960-hour audios for training. WER is evaluated on LibriSpeech test-clean and test-other test sets respectively. The test-clean contains simple and clean audios, test-other contains complex and noisy audios. All the neural networks are trained using Wenet toolkit~\cite{yao2021wenet}.
Models are trained with 3727 sub-word units, which generated from training transcripts of LibriSpeech using SentencePiece~\cite{kudo2018sentencepiece}.  Besides, the features embedding fed into the encoder is a  80-dimensional log filter-bank feature with 10ms frame shift and 25ms window size followed by Spec-Augment~\cite{park2019specaugment}. Two convolution layers are added with ReLU~\cite{agarap2018deep} activation function to downsample the frame rate to 4 with absolute position embedding layer.

The encoder has 12 layers, each conformer layer has convolution module with kernel size 31, model dimension 256 and two feed forward layers with dimension 2048. Dropout~\cite{srivastava2014dropout} rate is set to 0.1 for all layers. The decoder is a standard transformer decoder with 6 layers, each transformer layer has a FFN layer with dimension 2048, 4 attention heads and dropout is also set to 0.1 for all layers. 
CTC criterion is utilized to optimize the CTC loss as an auxiliary loss. Models are trained with the Adam~\cite{kingma2014adam} optimizer for about 150000 steps, peak learning rate is $4\times10^{-3}$ and warm up is 25000 steps. In inference, CTC greedy search and attention rescoring are applied to evaluate the performance of the model while beam size is set to 10. The amount of model parameter is about 49M.
Besides, We trained our models on one machine with 8 NVIDIA Tesla A100 GPUs. 

\subsection{Results}
All models are trained on LibriSpeech from scratch. The proposed model use kernel functions ReLU, Sigmoid~\cite{mount2011equivalence} and Exponential for comparison. We have observed that it's beneficial to change the number of attention head. Table~\ref{exp_ctcg} shows the result of CTC greedy search and attention rescoring respectively.

\begin{table}[th]
	\caption{WER(\%) of LBLA on LibriSpeech task using CTC greedy search decoding. B0 indicates the aforementioned Conformer baseline. Our LBLA model with the Sigmoid achieves better performance than the vanilla Conformer in every aspects.}
	\label{exp_ctcg}
    \scriptsize
    \centering
    \setlength{\tabcolsep}{1.5mm}{
    \begin{tabular}{lcccccc}
    \toprule
    \multirow{4}{*}{model} & \multirow{4}{*}{activation}&  \multirow{4}{*}{\makecell[c]{attention\\head}} &\multicolumn{2}{c}{test clean} & \multicolumn{2}{c}{test other} \\
    \cline{4-7}
    {}& {}& {}&\makecell[c]{ ctc\\ greedy\\search}&\makecell[c]{ attention\\ rescoring}&\makecell[c]{ ctc\\ greedy\\search}&\makecell[c]{ attention\\ rescoring}\\
    \midrule
    B0  & -  & 4 & 3.71  &3.2   & 9.67  & 8.74   \\
    \midrule
    R0  & ReLU     & 1 & 3.75         & 3.27         &10.15         & 9.16     \\
    R1  & ReLU     & 4 & 3.78         & 3.34         &9.81          & 8.87     \\
    R2  & ReLU     & 8 & 3.82         & 3.4          &9.97          & 8.98     \\
    \midrule
    E0  & Exponential      & 1 &{3.7}         & 3.28         & 9.68         & 8.88     \\
    E1  & Exponential      & 4 & 3.73         & 3.35         &{9.56}        & 8.82     \\
    E2  & Exponential      & 8 & 3.71         & 3.25         & 9.73         & {8.74}\\
    \midrule          
    S0  & Sigmoid  & 1 & 3.70         & 3.29         & 9.55         & 8.74     \\
    S1  & Sigmoid  & 4 & 3.70         & 3.27         & 9.59         & 8.83     \\
    S2  & Sigmoid  & 8 &\textbf{3.58} &\textbf{3.16} &\textbf{9.50} &\textbf{8.61}     \\
    \bottomrule
    \end{tabular}
    \vspace{-2.0em}
    }
    \end{table}
Firstly, we evaluate the performance of models that use CTC greedy search. As illustrated in Table~\ref{exp_ctcg}, Conformer baseline achieves a WER 3.71/9.67 on test-clean/test-other. Compared with baseline, $R$0-$R$2 perform slightly worse and achieve a relative degradation of WER by 1\%$\sim$2.9\% on test-clean and 1.4\%$\sim$4.9\% on test-other. The reason causing the performance degradation may be that $ReLU()$ forces negative value to be zero leading to loss some information of relevance weighted matrix. Compared with baseline, $E$0-$E$2 obtain the comparable performance . $E1$ even achieves a relative reduction of WER by 1.2\% on test-other. 
$S$0-$S$2 perform better and $S$2 achieves a relative reduction of WER by 3.5\% on test-clean and 1.5\% on test-other.
Experiments show that the proposed LBLA model achieves the comparable performance compared with vanilla Confomer, and exponential-based kernel function such as Sigmoid function and Exponential function could enlarge the distance among the proximity $\mathcal{P}(\cdot)$ to force the attention matrix to be more sparse.

We also evaluate the performance of models using attention rescoring. Conformer baseline achieves a WER 3.2/8.74 on test-clean/test-other. $R$0-$R$2 are slightly worse than baseline and achieve a relative degradation of WER by 2.1\%$\sim$6\% on test-clean and 1.5\%$\sim$4.8\% on test-other. 
$E$0-$E$2 achieve a relative degradation of WER by 2.5\%$\sim$4.7\% on test-clean and 0\%$\sim$1.6\% on test-other. However, $S$0-$S$2 obtain the best performance in our experiments and achieve a relative reduction of WER by 1.2\% on test-clean and 1.5\% on test-other. The proposed LBLA model also achieves better performance.

In our experiments, we obverse that validation loss is increasing while training last several epochs. It seems that the larger learning rate cannot make the LBLA model converge well in the late training stage. Consequently, We change the learning rate for late training stage while the rest strategies keep same by resume the model from middle of the training stage, all models' number of training steps keep same. As illustrated in Table~\ref{exp_finetune}. F1 outperforms the baseline trained from scratch by relatively 2.4\%$\sim$5.7\%.
\begin{table}[th]
	\caption{WER(\%) of LBLA with initialization on LibriSpeech task. Models half the learning rate for late training stage.}
	\label{exp_finetune}
    \footnotesize
    \centering
    \setlength{\tabcolsep}{1.7mm}{
    \begin{tabular}{lccccc}
    \toprule
    \multirow{4}{*}{model} & \multirow{4}{*}{initialize} & 
    \multicolumn{2}{c}{test clean} & \multicolumn{2}{c}{test other}\\
    \cline{3-6}
    {}& {} &\makecell[c]{ ctc\\ greedy\\search}&\makecell[c]{ attention\\ rescoring}&\makecell[c]{ ctc\\ greedy\\search}&\makecell[c]{ attention\\ rescoring}\\
    \midrule
    {Conformer} & scratch & 3.71 & 3.2 & 9.67 & 8.74     \\
    \midrule
    F0  & E0  & {3.6} & {3.19} & 9.62  & 8.78     \\
    F1  & S2  & \textbf{3.50} & \textbf{3.08} & \textbf{9.43}  & \textbf{8.53}     \\
    \bottomrule
    \end{tabular}
    }
    \vspace{-0.5em}
    \end{table}
\begin{table}[th]
	\caption{Inference speed of LBLA on LibriSpeech. It's defined as decoding utterance length per second. Three sub sets is used to evaluate the inference speed, which test-set1 contains all data in test-clean, test-set2 contains medium length speech longer than 10s in test-clean and test-set3 contains long speech longer than 20s in test-clean. All experiments are evaluated on a single thread of CPU.}
	\label{exp_speed}
    \footnotesize
    \centering
    \setlength{\tabcolsep}{5.0mm}{
    \begin{tabular}{lccc}
    \toprule
    model & test-set1 & {test-set2} & {test-set3}\\
    \midrule
    Conformer  & 22.8 & 22.0  & 20.7     \\
    \midrule
    R0  & 22.9 &23.6   & 23.0     \\
    R1  & \textbf{24.3} & \textbf{25.6}  & 25.2     \\
    R2  & 24.1 & 25.0   & 24.2     \\
    \midrule
    E0  & 23.7 & 24.3  & 23.8     \\
    E1  & 23.7 & 24.5  & 25.0     \\
    E2  & 24.0 & 24.8  & 25.0     \\
    \midrule
    S0  & 23.7 & 24.4  & 23.7     \\
    S1  & 23.8 & 24.7  & 24.1     \\
    S2  & 24.1 & 25.0  & \textbf{25.3}     \\
    \bottomrule
    \end{tabular}
    }
    \vspace{-1em}
    \end{table}
\begin{table}[th]
	\caption{Ablation study of LBLA, we remove its features: (1) removing cosine re-weight without additional relative positional embedding; (2) removing kernel function; (3) removing normalization for attention. NC represents no convergence.
	}
	\label{exp_ablation}
    \footnotesize
    \centering
    \setlength{\tabcolsep}{2.1mm}{
    \begin{tabular}{lcccc}
    \toprule
    \multirow{4}{*}{model} & 
    \multicolumn{2}{c}{test clean} & \multicolumn{2}{c}{test other}\\
    {} &\makecell[c]{ ctc\\ greedy\\search}&\makecell[c]{ attention\\ rescoring}&\makecell[c]{ ctc\\ greedy\\search}&\makecell[c]{ attention\\ rescoring}\\
    
    \midrule
    S2    &3.58 &3.16 &9.50 &8.61     \\
    \ \ $-$cosine re-weight    & {3.84} & {3.37} & 9.61  & 8.83     \\
    \ \ $-$Sigmoid    & {$NC$} & {$NC$} & $NC$  & $NC$     \\
    \ \ \ \ \ $+$ReLU   & 3.82  & 3.4  &9.97  & 8.98     \\
    \ \ $-$normalization    & {$NC$} & {$NC$} & $NC$  & $NC$     \\
    \bottomrule
    \end{tabular}
    \vspace{-2.0em}
    }
    \end{table}
    
To evaluate the time consumption of inference, we conduct the experiments on three test sets that contain different duration of speech. As illustrated in Table~\ref{exp_speed}, our proposed LBLA model obtains the better performance on time consumption. Compared with baseline, $R1$ increases the speed of inference by relatively 6\% on test-clean and $S2$ increases relatively 22\% on long speech sub set. Our proposed model will be more efficient with respect to the input sequence length.

LBLA model differs from a Conformer in some ways, specially, the inclusion of a kernel function and locality-biased re-weight mechanism. Normalization is applied similar to vanilla attention mechanism. We study these effects of these differences by removing features. Table~\ref{exp_ablation} shows the impact to the LBLA model. Kernel function and normalization are important to stabilize training until convergence. Locality-biased re-weight is an efficient way to obtain shift-invariant positional information and better performance.

\section{Conclusions}
\label{sec_con}
In this paper, we proposed a LBLA model which introduces a locality-biased linear mechanism in Conformer block to enjoy linear space-time complexities. The LBLA contains a kernel function to ensure linear computational complexity and a cosine reweighing matrix to impose more weight on neighbouring tokens. We evaluated it extensively on the LibriSpeech corpus. Compared with the vanilla Conformer, our proposed model enjoys linear computational complexity, obtains comparable WER performance and increases the inference speed by 6\%$\sim$22\%.
\bibliographystyle{IEEEtran}
\bibliography{template}
\end{document}